\documentstyle[aaspp,psfig]{article}

\def\etal{et al. }

\begin{document}

\title
{\bf Star Formation in Southern Seyfert 
Galaxies} 

\author{Duncan A. Forbes}
\affil{School of Physics and Astronomy, University of Birmingham,
Birmingham, B15 2TT}
\affil{Email: forbes@star.sr.bham.ac.uk}

\author{R. P. Norris}
\affil{Australia Telescope National Facility, CSIRO Radiophysics
Laboratory, P.O. Box 76,
Epping, NSW 2121, Australia}
\affil{Email: rnorris@atnf.csiro.au}

\begin{abstract}

We have produced radio maps, using the ATCA, 
of the central regions of six southern
Seyfert 2 galaxies (NGC 1365, 4945, 
6221, 6810, 7582, and Circinus) with circumnuclear 
star formation, to estimate the relative contribution of star 
formation activity compared to 
activity from the active galactic nucleus (AGN). The
radio morphologies range from extended diffuse structures to compact nuclear
emission, with no evidence, even in the relatively compact sources, 
for synchrotron self--absorption. 
In each case the radio to far--infrared (FIR) ratio has a value 
consistent with star formation, and in all 
but one case the radio to [FeII] ratio is also consistent 
with star formation. We derive supernova rates and conclude that, 
despite the 
presence of a Seyfert nucleus 
in these galaxies, the radio, FIR, and [FeII] line emission are
dominated by processes associated with the circumnuclear star formation
(i.e. supernova remnants and H~II regions) rather than with the AGN.

\end{abstract}

\keywords{galaxies: individual -- galaxies:
nuclei -- radio continuum: galaxies}

\section{Introduction}

Circumnuclear star formation is common in 
Seyfert galaxies, but the relationship between 
the Seyfert nucleus and the surrounding star 
formation is not well understood (see Filippenko 1992), and 
both evolutionary and causal relationships have been
suggested. For example, a nuclear 
starburst may develop into a massive star cluster or black hole 
(Norman \& Scoville 
1988), or conversely the energy output from an active galactic nucleus
(AGN) may 
trigger circumnuclear star formation (Sanders \& Bania 1976). 
More recently, studies have indicated that star formation is occurring in
and around the torus (e.g. Cid Fernandes \& Terlevich 1992; Davies \etal
1997) that is thought to dictate the type of Seyfert
nucleus we observe. 

Dust is a common feature of the circumnuclear environment of 
active galaxies. At optical wavelengths, it 
can obscure our view of the nucleus and hide any evidence of an AGN.  
However, radio observations are not only unaffected by dust but also have
the advantage of high spatial resolution, 
and can be important in understanding 
the role of the various processes in active galaxies. 
For example, starburst galaxies generally have diffuse radio morphologies
dominated by synchrotron emission from cosmic rays accelerated by supernovae,
while
Seyfert galaxies sometimes exhibit 
well--collimated radio jets and a compact nuclear radio source
(see review by Condon 1992).  

High-resolution radio synthesis images, with sub-arcsec 
resolution,  have found
compact sources at the nucleus, and 
extended emission consisting of 
radio jets, ouflows, diffuse emission, 
and discrete sources such as HII regions and SNRs (e.g.  
Ulvestad \& Wilson 1984; Kronberg \etal 1985; Antonucci \& Ulvestad 1988; 
Carral, Turner \& Ho 1990; Condon \etal 1991; 
Forbes \etal 1994; Collison \etal 1994; Sandqvist \etal 1995).  

High spatial resolution, multi--frequency radio observations can 
also be used to
test the idea of advection-dominated accretion flows around black
holes. For such flows the radio emission depends strongly on the mass of
the black hole and is characterised by an inverted radio spectrum (Yi \&
Boughn 1997). 

Here we present 3 and 6cm 
radio continuum images from the Australia Telescope Compact Array 
(ATCA) of 6 such 
galaxies which
show evidence for narrow high excitation optical forbidden lines, and are 
classified as Seyfert 2s. In the case of NGC 1365, broad lines have
also been detected from the nuclear region.
All six galaxies show evidence for circumnuclear star formation, and several
are barred. 
We discuss the radio 
morphologies of these galaxies and possible emission mechanisms, and we
compare the radio data 
with that from other wavelengths to assess the relative contributions 
to the radio flux from star formation and the AGN.

\section{Observations and Data Reduction}

The observations were carried out using the Australia Telescope Compact 
Array at 4.79 GHz (6 cm) and 8.64 GHz (3 cm) 
simultaneously. The observational parameters are given in Table 
1. The same amplitude calibrator (1934--638) was observed with each 
galaxy, with an assumed flux density of 5.83 Jy 
at 6 cm and 2.84 
Jy at 3 cm. Observations of the target galaxy and a phase 
calibrator were alternated throughout the observing run, with a total 
of about 
10 hours spent on-source on each galaxy.  
The data were edited, calibrated and CLEANed using the AIPS 
software package. The typical half--power beam--width
(HPBW) of the final images are 2 arcsec at 6 cm and 1 arcsec at 3 cm. 

Because these observations and their analysis were optimised for 
studying the nuclear region, 
they are relatively insensitive to extended emission. Thus our images 
are primarily of the 
nuclear region, and we will therefore not show any smooth extended structure 
on scales of tens of arcsec.

\section{Results and Discussion}

In Figures 1 to 12 we show the 3 and 6cm radio images for the six 
galaxies in our sample. These galaxies do not represent a complete sample
in any sense, but rather were chosen as relatively well--known southern
Seyfert 2 galaxies that generally lacked high resolution radio maps. 
Our radio data for the Circinus galaxy have been presented 
elsewhere along with near--infrared line imaging (Davies \etal 1997),
although we include it here for comparison purposes with the other
galaxies.  The images show a variety of
radio morphologies which range from Circinus, with a strong,
compact nucleus, to NGC 1365, with a extended region of discrete sources
or hot-spots. The beam shape is shown in the lower left of each image. 

Care is needed when interpreting these images, as (a) our observations 
are optimised for 
studying the nuclear region, 
and so our images will not correctly represent the extended 
diffuse emission from the disk of the galaxy, and (b) most of the 
galaxies have high 
inclinations so that 
confusion effects may play some role in the observed radio morphology. 

Flux measurements in 2 arcsec and 6 arcsec diameter apertures for each
galaxy are given in Table 2. We also list the 6cm to 3cm spectral index 
after smoothing the 3cm image to match the 6cm resolution.
The spectral indices (F $\propto \nu ^{\alpha}$) range from flat ($\alpha$
$\sim$ 0) to steep ($\alpha$ $\sim$ --1). The total 6cm flux in Table 2
is given both for our images and for the single--dish observations 
by Wright \etal (1994, 
1996). The difference between these indicates the amount of  diffuse 
emission which is 
missing from our images. We also calculated the 
maximum brightness temperatures for each image (both 3 and 6 cm)
and the maximum value is given for each source.

In Table 3 we list various derived quantities for the sample including
Hubble type, distance, inclination, 6cm and [FeII] line luminosities and 
the SN rate.
The [FeII] measurements are from Moorwood \& Oliva
(1988) in a 6 arcsec diameter aperture, except for the Circinus galaxy 
in which
we use a nuclear [FeII] flux from Davies \etal (1997). None of the [FeII]
fluxes have been corrected for extinction.
The SN rate is calculated separately from both the 6cm and [FeII] line
flux in the 6 arcsec aperture (corresponding to $\sim$ 500 pc at a typical
distance of 20 Mpc). 
The 6cm SN rate is calculated assuming that all of the 6cm flux is
non--thermal emission from cosmic rays accelerated by SNRs 
(e.g. Condon \& Yin 1990). 
This may give an overestimate of the rate
because it ignores any nuclear flux (which may be significant) 
and the contribution from thermal
emission (which is unlikely to be significant). For the [FeII] SN rate 
we simply assume an
average luminosity of 2 $\times$ 10$^{36}$ erg s$^{-1}$ over an adiabatic
lifetime of 2 $\times$ 10$^{4}$ yrs (e.g. Norris \& Forbes 1995). For most
galaxies the two methods give rates within a factor of two, the notable
exception being NGC 4945 (which is discussed further below).

\subsection{Radio Spectral Indices}

The radio
spectral index of Seyfert and starburst galaxies is determined by 
four mechanisms.

1. Cosmic rays within the galaxy are generated and
re--accelerated by supernovae and supernova remnants, and then interact with
the interstellar magnetic field to emit  
synchrotron emission. This synchrotron emission, with a typical 
spectral index of $\alpha$ $\sim$
--0.7, is expected to dominate the radio power of starburst
galaxies, and should appear as a 
diffuse component in radio images of these sources.
 
2. Relativistic particles ejected from the massive black hole at 
the nucleus of a galaxy 
may generate  intense synchrotron emission, similar to that seen 
in radio--loud galaxies 
and quasars. The cores in these radio-loud objects typically have a 
flat--spectrum core, 
indicating synchrotron
self--absorption, and steep--spectrum extended radio--lobes, suggesting
cooling of high--energy electrons. However, synchrotron self--absorption is 
important only
for brightness temperatures greater than 10$^{10}$ K (Condon 1992). 
Most Seyfert galaxies, 
on the other hand, are observed to have much lower brightness 
temperatures than this in the core,
so that synchrotron self-absorption is not significant in these 
sources. This is confirmed by the observed
core spectral indices, which are frequently in the region of --0.7. 

3. When the radiative efficiency in the accretion disk is low, an 
advection-dominated accretion flow (ADAF) may 
operate. The radio emission in these sources is 
dominated by synchrotron emission from a hot plasma, 
and the emission from such flows is 
predicted to give rise to inverted spectra with
typical indices of +0.4 (Yi \& Boughn 1997). 
The ADAF radio emission mechanism has only 
recently been proposed and should be 
regarded as untested at this stage. 
The data here are unlikely to provide a 
definitive test because of 
insufficient resolution, and all radio spectral indices 
measured here are negative. We will therefore not 
consider this mechanism further, except to  
note that this mechanism would be indicated by 
inverted-spectrum emission from a 
low-brightness-temperature core. No source here falls into this category.

4. H II regions in our
galaxy generate free--free emission from hot electrons. Most are
optically thin, giving a flat spectrum, although some compact H II regions
become optically thick at centimetre wavelengths, giving a spectral index
$\sim$+2. However, the integrated flux of 
such regions is generally insignificant
compared to the synchrotron emission of the host galaxy.

5.  The radio emission from ultra--luminous infrared galaxies is optically
thick to free--free absorption, so that the typical synchrotron spectrum
of these galaxies is flattened at low frequencies (Condon et al. 1991).

The combined result of these effects in Seyfert and starburst galaxies is to
produce a typical radio spectral index of --0.7 
(from the extended synchrotron emission)
with a flattening at low
frequencies in some starburst sources because of free--free absorption. 

Table 2 shows that the nuclear spectral indices of three of the 
galaxies (NGC 1365, 
NGC 6221, NGC 7582) is --0.5 or steeper on both the 
2 arcsec and the 6 arcsec scale, showing evidence 
for neither free--free nor synchrotron absorption. 
In the other three galaxies, the cores have 
flatter spectra, but the brightness temperatures ($\le$ 8300 K) 
are too low for 
synchrotron self--absorption, 
indicating that free--free 
absorption is responsible for the flattening. 

Of course, we cannot rule out the presence of a weak 
synchrotron self--absorbed core in the nucleus of any of 
these galaxies. However, comparison of the radio fluxes in a 
2--arcsec aperture with the flux in a 
6--arcsec aperture in Table 2 shows that the luminosity of any 
such core is small compared to the surrounding emission. 
Therefore any such core does not contribute significantly 
to the overall energy budget of the 
nuclear region of the galaxy, and is not 
responsible for the overall flat spectrum of the nuclear region..

This degree of free-free absorption flattening indicates either a 
high star formation rate (Condon \etal 1991) or that we are 
viewing the AGN through optically--thick 
narrow-line-region clouds (Roy \etal 1994). 

\subsection{The Radio - [FeII] Correlation}

Forbes \& Ward (1993) discovered that the 6cm radio emission in the
central regions of active galaxies is strongly correlated with the
near--infrared [FeII] 1.64$\mu$m 
line emission. This relation exists over several
orders of magnitude. With a larger sample, Simpson \etal (1996) were
able to show that Seyfert and starburst galaxies follow slightly
different radio--[FeII] relations. For starburst galaxies the relation,
with slope $\sim$ 1, 
can be reasonably explained by SNRs which are responsible for both the
non--thermal radio emission and the fast shocks that provide the
[FeII] excitation. However, the situation for Seyfert galaxies (which
reveal a correlation slope of $\sim$ 0.7) is less
clear. Simpson \etal argued that photo-ionisation from the Seyfert
nucleus can cause this relationship, with a contribution from radio--jet
induced shocks in some cases. 

In Fig. 13 we show the [FeII]/6cm ratio for our sample galaxies,
compared with the Seyfert and starburst relations of Simpson
\etal (1996). The 1$\sigma$ 
dispersion of galaxies about the relations is $\sim$ $10^{0.5}$.
For the Seyferts studied here, we find a
large degree of star--formation activity compared to 
photo--ionisation from an AGN, and so we might expect them to lie closer to
the starburst relation than the Seyfert one. This indeed appears to be the
case for four galaxies, although one (NGC 7582) is 
closer to the Seyfert relation
and NGC 4945 falls well away from either relation. The  
[FeII]/6cm ratio of NGC 4945 is about a factor of 100 lower 
than typical active 
galaxies, and we discuss this
further in Section 3.4 below.
We note however that given the dispersion
in the relations, and the low luminosities ($\le$ 10$^{40}$ 
erg s$^{-1}$) of the galaxies studied here, 
this is not a sensitive test of the 
excitation mechanism. 

\subsection {The Radio -- FIR Correlation}

Normal spiral and starburst galaxies show a 
tight correlation between their radio and 
FIR luminosity (e.g. Wunderlich et al. 1987). This correlation, 
which extends over 
five orders of magnitude, is true for both flux density and luminosity, 
and cannot be 
accounted for by selection effects, or by a simple ``richness effect''. 
While a detailed 
mechanism to explain this correlation 
has yet to be established, it is almost certainly 
the result of star formation, which generates both the synchrotron radio 
emission and 
the thermal FIR emission. This is supported by the fact that all 
objects that are 
dominated by star formation (HII galaxies, normal spirals, 
starburst galaxies) do 
follow the correlation.

On the other hand, Sopp \& Alexander (1991) showed that 
radio--loud quasars and 
radio galaxies clearly do not follow the radio--FIR correlation.
Thus whether or 
not a galaxy follows this correlation may be used as an indicator of 
the dominant 
radio luminosity source of the galaxy.

Norris \etal (1988) and Roy \etal (1997) showed that 
Seyfert galaxies, unlike radio--loud  
quasars, do roughly follow the radio--FIR correlation, 
but with a looser fit than normal 
spirals and starbursts. This suggests that the bolometric 
luminosity of Seyfert galaxies 
may be dominated by star formation. This is supported by off--nuclear 
optical and 
infrared observations of Seyferts, which show the same line ratios and 
luminosities as 
starburst galaxies (Bransford et al. 1997). Thus, although the nuclear 
optical spectra of 
Seyfert galaxies are clearly dominated by an AGN,
in many cases the 
integrated radio emission and the FIR emission are dominated 
not by the AGN 
but by circumnuclear star formation.

The degree to which an individual galaxy 
follows this correlation is most 
conveniently expressed by the parameter q -- 
the logarithm of the FIR to radio ratio. 
The conventional definition of q follows that of Helou \etal 
(1985), who define q in terms of the 1.49 GHz radio flux. 
For our purposes, we adapt Helou's 
definition to our observing frequency of 
4.8 GHz by assuming a spectral index of --0.7, and therefore define it as

\hspace{.5in}q $\equiv$ log[(FIR/3.75 x 10$^{12}$ Hz)/(2.26 x S$_{\rm 4.8
GHz}$]\hspace{1.5in}(1)

\hspace{.5in}where FIR $\equiv$ 1.26 x 10$^{-14}$(2.58S$_{\rm
60\mu}$+S$_{\rm 100\mu}$)\hspace{2in}(2) 

Typical values of q from the IRAS Bright Galaxy Sample are 2.34 for normal
spirals, 2.21 for starburst galaxies, and less than 2 for radio--loud AGNs 
(Condon \etal 1991). 

All the galaxies studied here except NGC 4945 
have q in the range 2.2 to 2.3, 
which places them firmly in the middle of the 
radio--FIR correlation, 
and suggests that most of their 
radio and FIR luminosity is produced by star formation.
We discuss the case of NGC 4945 (q = 1.88) below.

\subsection{Individual Galaxies}

Here we discuss each galaxy in turn, starting 
with an extended discussion of NGC 4945. To 
avoid repetition, we note that in every case other than NGC 4945, 
the spectral index, radio--FIR ratio, and [FeII]--radio 
ratio are all consistent with star formation, rather than an AGN, 
being the dominant source of radio emission.

\noindent
{\bf NGC 4945} This infrared luminous galaxy is nearly
edge--on and is located in a nearby loose group. Although we list it as a
barred galaxy in Table 3, there is a continuing debate about the 
reality of the bar
(e.g. Harnett \etal 1989). 
Koornneef (1993) described NGC 4945 as a post--starburst 
galaxy with no evidence for an AGN. However Moorwood \& Oliva (1994) have
argued that the central regions do show signs of ongoing young star
formation. Evidence for a heavily obscured AGN now come from the 
variable hard X--rays (Iwasawa \etal 1993), and the presence of a 
compact radio core in VLBI observations (Sadler \etal 1995).
The galaxy contains a thick torus or ring  with a radius of $\sim$ 150 pc 
(Koornneef 1993; Moorwood \etal 1996). 
Harnett \etal (1989) found that the radio emission has a strong central
contribution with  emission extended 10 arcmin
perpendicular to the major axis.
Multi--frequency observations have been carried 
out by Elmouttie \etal (1997). They focused on the large scale structure 
using a beam size of $\sim$ 25$^{''}$, and found that the spectral index 
steepens from the central region to the main disk of the galaxy. 

Furthermore, NGC 4945 is notable for the fact that it is one of the 
few galaxies (along with Circinus) known to contain water 
megamasers. Such megamasers have been cited 
in NGC 4258 (Miyoshi \etal  1995) as the  
strongest evidence known for a massive black hole in an AGN. 
Preliminary VLBI imaging (Greenhill \etal 1997) of the 
megamasers supports 
the model that they are in a Keplerian disk 
surrounding the black hole. 

Our radio image is dominated by strong nuclear emission, and 
emission extended
along the disk of the galaxy. However there is also 
evidence of some filamentary structure perpendicular to the major
axis. Such extended emission may be associated with the
outflowing superwind in this galaxy (Nakai \etal 1989; Lipari, Tsvetanov \& 
Macchetto 1997). 
The extended emission has a steeper spectral index 
($\alpha \sim -0.8$) than the radio nucleus, however we note that our data 
are less sensitive to extended emission (particularly at 3cm) which makes the 
spectral index somewhat less certain.  
The nucleus has a relatively flat spectral index of $\alpha$ =
--0.3. 
The brightness temperature in the central 2 arcsec of this source
(i.e. 7000 K) 
is still far too low for synchrotron 
self--absorption, indicating that the star formation activity 
is particularly intense, to provide the necessary free--free absorption. 
The obscuration inferred from the
X--ray data suggest that the extinction 
towards the AGN could be as high as A$_V$ $\sim$ 2500. 

We noted above (in Section 3.3) that the radio--FIR ratio for 
this galaxy is unusually low (i.e. q = 1.88), which at first sight appears
to suggest that 
an AGN is responsible for much of the radio emission. 
However, this galaxy is so near that 
not all the FIR flux was in the single IRAS aperture, and so the 
IRAS flux listed in the IRAS 
Point Source Catalog may be an underestimate. Rice \etal (1988) 
have estimated the total 
FIR flux by co--adding IRAS images and obtain a higher value for the FIR 
fluxes, which raises 
the value of q to 2.1, suggesting that the radio emission in this galaxy 
is again dominated by 
star formation rather than by an AGN.

An interesting property of NGC 4945 is that the ratio of the [FeII] line
luminosity to 6cm radio emission is only 0.63, which is almost a factor of
100 less than is 
typical for active galaxies (see Fig. 13). We now consider a number of 
possible reasons for this.

\begin{itemize} 

\item The reduced ratio could be due to extinction (by dust) 
of the [FeII]. However,  Moorwood \&
Oliva (1994) estimate that the extinction in the [FeII] line emitting zone
is 1.8 mag or a factor of five, which is insufficient to produce the 
observed effect.

\item It could be because of nuclear radio emission from an AGN which 
is not accompanied by [FeII] line emission.  
We have shown above that the radio/FIR ratio 
for the galaxy as a whole is consistent with 
star formation activity. However, the central 6 arcsec (over which 
we measure the [FeII]/6cm 
ratio) contributes only 
9\% of the total radio flux, and we have no information on 
the radio/FIR ratio in the nucleus, 
so the radio flux from the AGN could be 
abnormally large. In this case, we would expect the [FeII]/6cm 
ratio to approach the 
usual value as we increase the area over which we integrate the flux 
(for both 6 cm 
and [FeII]). However, Moorwood \& Oliva
(1994) quote a total [FeII] flux over an emitting region of 18 
$\times$ 21 arcsec to be 12 $\times$ 10$^{-14}$ erg s$^{-1}$ cm$^{-2}$, 
or an
observed log luminosity of 38.81 erg s$^{-1}$. 
The 6cm radio luminosity over a similar area is 40.21 erg 
s$^{-1}$, giving
a [FeII]/6cm ratio of 0.03, 
which is even lower than the value in the nucleus, indicating 
that the [FeII]/6cm ratio falls off with
distance from the galaxy centre, and that the low value is not 
a consequence of nuclear radio emission. 

\end {itemize}

Thus the abnormally low 
[FeII]/6cm ratio in NGC 4945 of 0.63 is produced in the region
surrounding the nucleus, where the 
radio emission (with a spectral index of $\sim$ --0.7) is
due to SNRs in the galaxy disk, the outflowing starburst
superwind discussed above, or perhaps a radio jet. 
Pure SNRs produce ratios of about 500 
i.e. well in excess of typical
galaxy values, so this would tend to give an enhanced ratio. In a $6^{''}
\times 6^{''}$ aperture, the superwind galaxies 
M82 and NGC 253 have [FeII]/6cm ratios of 75 and 52
respectively, although the superwind itself in NGC 253 does not seem to 
produce significant [FeII] line emission (Forbes \etal 1993). Again such
ratios are significantly higher than seen in NGC 4945. 
The data for Seyfert galaxies with clear radio jets are limited. For NGC
4151 and NGC 1068 
the measured ratios are 28 and 9. This is closer to the
NGC 4945 value but still a factor of at least 10 too high. 
We conclude that the abnormal
[FeII]/6cm ratio in NGC 4945 is 
due to either (a) a starburst superwind, 
which produces substantial radio emission but 
very little [FeII] line flux (due perhaps to an unknown 
excitation effect or low density in the wind), 
or (b) a radio jet which dominates the extended radio emission 
on the few--arcsec scale 
but which does not produce significant [FeII] emission. 

\noindent
{\bf NGC 1365} This is a well--studied barred galaxy. The central 
region reveals broad and narrow emission lines 
(Veron \etal 1980) surrounded by a 
circumnuclear ring of star formation (Edmunds \& Pagel 1982; 
Saikia \etal 1994). 
The star formation, combined with the obscuring effects of dust, give the 
appearance of a hot-spot nucleus (Sersic \& Pastoriza 1965). A high 
excitation outflow 
from the nucleus has been seen (e.g. Hjelm \& Lindblad 1996). 

High resolution radio continuum observations have been reported by
several workers (e.g. Sandqvist, Jorsater \& Lindblad 1982, 1995). In
particular,  Sandqvist, Jorsater \& Lindblad (1995) observed it 
with the VLA at  20, 6, and 2 cm.
Their radio images revealed a weak nucleus surrounded by a elongated 
$\sim$ 8 $\times$ 20 arcsec (a/b = 0.4) ring of hot-spots, or components.
They labelled a number of components A--H, of which B is blended with A
and C is blended with D at $\ge$ 1 arcsec resolutions. 
Our radio image, shown in Fig. 1, is consistent with theirs, 
except that we identify
one additional component to the SW, which we call `J'. The hot-spots
generally have steep spectra with 6cm luminosities of 
$\sim$ 10$^{36}$ erg s$^{-1}$ which suggests that the radio emission from 
each component is made up of several SNRs.   
A combined radio and X--ray study of the nucleus and surrounding regions 
has been carried out by Stevens, Forbes \& Norris (1998). 
The radio nucleus does not appear to have an X--ray counterpart. 
Furthermore the X--ray emission is consistent with star formation processes. 
Stevens \etal 
conclude that if NGC 1365 harbours a black hole it is largely inactive. 

\noindent
{\bf NGC 6221} Located in a small group, NGC 6221, is a barred galaxy
with a weak Seyfert nucleus. The galaxy may be interacting with NGC
6215 and has a nuclear bar (Koribalski 1996). 
The radio emission from NGC 6221 is extended in an symmetric 
bar--like structure. The spectral index of the nucleus and bar are 
non--thermal with $\alpha$ $\sim$ --0.6, indicative of SNRs. 
The radio morphology
and other properties are all consistent with star formation being the 
dominant source of radio 
emission.

\noindent
{\bf NGC 6810} This early type spiral is the most distant in our sample and
has not been well studied to
date. It may contain a bar and ring structure (Buta 1995), and does 
not appear to have been
imaged before at radio wavelengths. Our radio
image reveals a dominant nucleus surrounded by diffuse extended radio
emission. Both the nucleus and surrounding region have flat spectral
indices, but the brightness temperature 
is too low for this to be attributable to
synchrotron self--absorption, 
which suggests that the radio spectrum is flattened 
by free--free absorption from young star formation. The radio morphology and 
other radio properties are 
all consistent 
with star formation being the dominant source of radio emission. 
Interestingly, recent 
high resolution optical spectra do not confirm the status of 
NGC 6810 as a Seyfert galaxy (Heisler 1998), thus it appears to 
have been misclassified.

\noindent
{\bf NGC 7582} This narrow line X--ray 
galaxy is located in the Grus loose group along
with NGC 7590 (a Seyfert 2; Ward \etal 1980), NGC 7552 (a starburst;
Forbes \etal 1994) and NGC 7599. Several HI bridges connect group members
(Koribalski 1996). Morris \etal (1985) provide evidence for both a
rapidly--rotating $\sim$ 1 kpc ring of circumnuclear star formation and
high excitation gas moving outwards from the nucleus. 

Ulvestad \& Wilson (1984) imaged the galaxy at 6 cm using 
the VLA with a beam size of $\sim$ 1.5 arcsec. They measured a total
6cm flux of 69 mJy. We find a linear, double--peaked morphology to the radio
emission. The southern peak appears to lie at the centre of the outer
radio isophotes and is presumably the true nucleus. The nucleus has a steep
spectral index ($\alpha$ = --0.7) indicating non--thermal emission. 
To the NW by
$\sim$ 3 arcsec lies a second peak, which could be a second
nucleus. However, it lies roughly along the bar/major axis position
angle (P.A. $\sim$ 150$^{\circ}$). It has a 2 arcsec diameter 6cm flux of
10 mJy and a spectral index of --0.9.
This second peak could
therefore be a radio jet or simply a discrete star formation region 
occurring along the galaxy bar. The inferred SN rate in the central
6 arcsec (870 pc) is the highest in our sample (except possibly for NGC
4945) at about 1 SN every 8 years.
Although the  radio morphology 
suggests a linear Seyfert jet, the spectral index, radio--FIR 
ratio, and [FeII]--radio ratio are all consistent with 
star formation being the dominant source of 
radio emission.

\noindent
{\bf Circinus} The Circinus galaxy is perhaps the closest Seyfert
galaxy known but is difficult to observe due to its proximity to 
the Galactic 
plane and large internal obscuration. Confirmation of an AGN comes from the
the presence of high excitation coronal lines (Oliva \etal 1994), X--ray
emission (Matt \etal 1996), and a compact radio core (Heisler \etal 1998).

Like NGC 4945, Circinus is one of the 
few water 
megamaser galaxies.
The megamasers in Circinus 
are stronger but less extreme than those in NGC 4258, 
and have the curious property of 
fluctuating on a time scale of minutes 
(Greenhill \etal 1997), indicating a particularly 
compact source. Preliminary VLBI imaging of the 
megamasers (Ellingsen \etal 1998) 
indicates that the maser region is 
extended with a velocity gradient aligned with that of the 
parent galaxy, and perpendicular to the jet. 
We regard this as strong evidence for a massive black hole 
in this galaxy.

Marconi \etal (1994) found 
both a circumnuclear starburst ring and an ionisation 
cone. The [OIII] ionisation cone is asymmetric extending only to 
the NW, with 
some high excitation lines also seen in the cone region. They estimated the 
extinction to the nucleus to be A$_V$ $\sim$ 20. 
The HI gas distribution shows spiral arms, a bar and a 
central `hole' (Koribalski 1996, Jones \etal 1998). 
High resolution observations of the 
central region indicate a rapidly--rotating gas 
ring with a diameter of $\sim$ 
400 pc (Koribalski 1996). 
Observations with the ATCA have been carried 
out at 13 and 20 cm by Elmouttie \etal (1995). 
They found extended radio lobes perpendicular 
to the galaxy major axis (position angle = 30$^{\circ}$) with a
spectral index of $\alpha \sim$ --0.7. 

Our 3 and 6cm radio images, 
observed as part of this project, have been
published, along with near--infrared line images, by Davies \etal (1997). 
The radio data indicate that the nucleus is marginally resolved with a flat
spectral index. The low brightness temperature indicates that
this is due to free--free
absorption rather than synchrotron self--absorption in a compact AGN source. 
There are also 
faint hints of extended emission which may be
associated with outflowing material. Despite the clear indication of 
a compact AGN
in the radio images, the other radio indicators are consistent with 
the more extended 
radio emission being dominated by star formation activity.\\

\section{Summary and Conclusions}

We have imaged six southern active galaxies at both 3 and 6 cm using the
ATCA. The radio emission reveals a variety of morphologies, from 
compact nuclear emission to extended diffuse structures. Several galaxies
show radio components and linear features from the nucleus. 
Four galaxies (NGC 4945, 6810, 7582 and Circinus) 
reveal strong, fairly compact nuclear radio sources which may be due to
synchrotron emission associated with an AGN, although the nuclear sources
are not sufficiently compact for synchrotron self--absorption to be important. 
Three galaxies (NGC 1365, 7582 and Circinus) contain 
hints of jet--like features, but none are very convincing. 
NGC 1365 contains a very weak, 
compact nuclear radio source for which there is little evidence at radio or 
X--ray (Stevens \etal 1998) wavelengths for an AGN. 
NGC 6221 shows no evidence in the radio for the 
presence of an active nucleus, and in the case of NGC 6810 there are doubts 
about its optical classification as a Seyfert (Heisler 1998). 

For most of the galaxies the SN rate, in the central $\sim$ 500 pc, 
inferred from [FeII] line luminosities
are similar to those from 6cm luminosities, suggesting that the extended 
non--thermal radio emission is dominated by diffuse synchrotron 
emission from cosmic rays accelerated by SNRs. In NGC 1365, the
circumnuclear radio components are dominated by collections of SNRs which are
more intense at radio wavelengths than the nucleus itself. 
For NGC 4945 additional radio emission appears to be coming from an 
outflowing superwind, perpendicular to the galaxy disk.

In all six galaxies the radio emission is very much dominated by processes
associated with star formation. The AGN, if present, makes only a small
energetic contribution, and we are unable to detect the presence of any 
advection-dominated accretion flows. 

\noindent{\bf Acknowledgments}\\
The Australia Telescope is funded by the
Commonwealth of Australia for operation as a National Facility managed
by CSIRO. We thank B. Koribalski and I. Stevens for their comments on the 
manuscript.  We also thank the referee T. Muxlow for several useful
suggestions. \\

\noindent{\bf References}\\
Antonucci, R. R. J., Ulvestad, J. S., 1988, ApJ, 330, L97\\
Bransford, M.A., Appleton, P.N, Heisler, C.A.,  
Norris, R.P., Marston, A.P., 1997, ApJ, submitted\\
Buta, R., 1995, ApJS, 96, 39\\
Carral, P., Turner, J., Ho, P. 1990, ApJ, 362, 434\\
Collison, P. M., \etal 1994, MNRAS, 268, 203\\
Cid Fernandes, R., Terlevich, R., 1992, Relationships between Active
Galactic Nuclei and Starburst Galaxies, ed. A. V. Filippenko, ASP: San
Francisco, p. 241\\
Condon, J. J., 1992, ARAA, 30, 575\\
Condon, J. J., Yin, Q. F., 1990, ApJ, 357, 97\\
Condon, J. J., Huang, Z.P., Yin, Q.F., Thuan, T.X., 1991,  ApJ, 378, 65\\
Davies, R. I., \etal 1997, MNRAS, in press\\
Edmunds, M. G., Pagel, B. E. J., 1982, MNRAS, 198, 1089\\
Ellingsen, S., \etal 1998, in preparation\\
Elmouttie, M., \etal 1995, MNRAS, 275, 53\\
Elmouttie, M., \etal 1997, MNRAS, 284, 830\\
Filippenko, A. V., 1992, Relationships between Active
Galactic Nuclei and Starburst Galaxies, ed. A. V. Filippenko, ASP: San
Francisco\\
Forbes, D. A., Ward, M. J., 1993, ApJ, 416, 150\\
Forbes, D. A., Norris, R. P., Williger, G. M., Smith,
R. C., 1994, AJ, 107, 984\\
Forbes, D. A., \etal 1993, ApJ, 406, L11\\
Greenhill, L. J., Moran, J. M., Herrnstein, J. R., 1997, ApJ, 481, 23\\
Harnett, J. J., Haynes, R. F., Klein, U., Wielebinski, R., 1989, A \& A,
216, 39\\
Heisler, C. 1998, personal communication\\
Heisler, C., \etal 1998, in preparation\\
Helou, G., Soifer, B. T., Rowan-Robinson, M., 1985,
ApJ, 298, L7.  \\
Hjelm, M., Lindblad, P. O., 1996, A \& A, 305, 727\\
Iwasawa, K., \etal 1993, ApJ, 409, 155\\
Jones, K., \etal 1998, MNRAS, in press\\
Koorneff, J., 1993, ApJ, 403, 581\\
Koribalski, B., 1996, Minnesota Lectures on Extragalactic Neutral Hydrogen
ed. E. Skillman, ASP Conf. Series 106, p 238\\
Kornberg, P. P., Biermann, P., Schwab, F. R., 1985, ApJ, 291, 693\\
Lipari, S., Tsvetanov, Z., \& Macchetto, F. 1997, ApJS, 111, 369\\
Marconi, A., Moorwood, A. F. M., Origlia, L., Oliva, E., 1994, The
Messenger, 78, 20\\
Matt, G., \etal 1996, MNRAS, 281, 69\\
Miyoshi, M., Moran, J., Herrnstein, J., Greenhill, L., Nakai, N., Diamond,
P., Inoue, M., 1995, Nature, 373, 127\\
Moorwood, A. F. M., Oliva, E., 1988, A \& A, 203, 278\\
Moorwood, A. F. M., Oliva, E., 1994, ApJ, 429, 602\\
Moorwood, A. F. M., van der Werf, P. P., Kotilainen, J., Marconi, A., 
Oliva, E., 1996, A \& A, 308, 1\\
Morris, S. L., \etal 1985, MNRAS, 216, 193\\
Nakai, N. 1989, PASJ, 41, 1107\\
Norman, C., Scoville, N., 1988, ApJ, 332, 124\\
Norris, R. P., Forbes, D. A., 1995, ApJ, 446, 594\\
Norris, R. P., Allen, D. A., Roche, P. F., 1988,
MNRAS, 234, 773\\
Oliva, E., \etal 1994, A \& A, 288, 457\\
Rice, W., Lonsdale, C.J., Soifer, B.T., Neugebauer, G., Kopan, E.L., 
Lloyd, A.L., de Jong, T., \& Habing, H.J., 1988, ApJS, 68, 91.\\
Roy, A. L., Norris, R. P., Kesteven, M. J., Troup, E. R.,
Reynolds, J. E., 1997, MNRAS, submitted\\
Sadler, E., \etal 1995, MNRAS, 276, 1373\\
Saikia, D. J., Pedlar, A., Unger, S. W., Axon, D. J., 1994, MNRAS, 270, 465\\
Sanders, R. H., Bania, T. M., 1976, ApJ, 204, 341\\
Sandqvist, A., Jorsater, S., Lindblad, P. O., 1982, A \& A, 110, 336\\
Sandqvist, A., Jorsater, S., Lindblad, P. O., 1995, A \& A, 295, 585\\
Sersic, J. L., Pastoriza, M., 1965, PASP, 77, 287\\
Simpson, C., Forbes, D. A., Baker, A. C., Ward, M. J., 1996, MNRAS 
283, 777\\
Sopp, H. M., Alexander, P. 1991, MNRAS, 251, 14P\\
Stevens, I. R., Forbes, D. A., Norris, R. P. 1998, in preparation\\
Ulvestad, J. S., Wilson, A. S., 1984, ApJ, 278, 544\\
Ulvestad, J. S., Wilson, A. S., 1984, ApJ, 285, 439\\
Veron, P., \etal 1980, A \& A, 87, 245\\
Ward, M. J., Penston, M. V., Blades, J. C., Turtle, A. J., 1980, MNRAS,
193, 563\\
Wright, A. E., Griffith, M. R., Burke, B. F., Ekers, R. D., 1994, ApJS, 91,
111\\
Wright, A. E., Griffith, M. R., Hunt, A. J., Troup, E., Burke, B. F.,
Ekers, R. D., 1996, ApJS, 103, 145\\
Wunderlich, E., Klein, U., Wielebinski, R., 1987, 
A \& AS, 69, 487\\
E. Skillman\\
Yi, I., Boughn, P., 1997, preprint (astro-ph/9710147) \\

\begin{figure*}[p]
\centerline{\psfig{figure=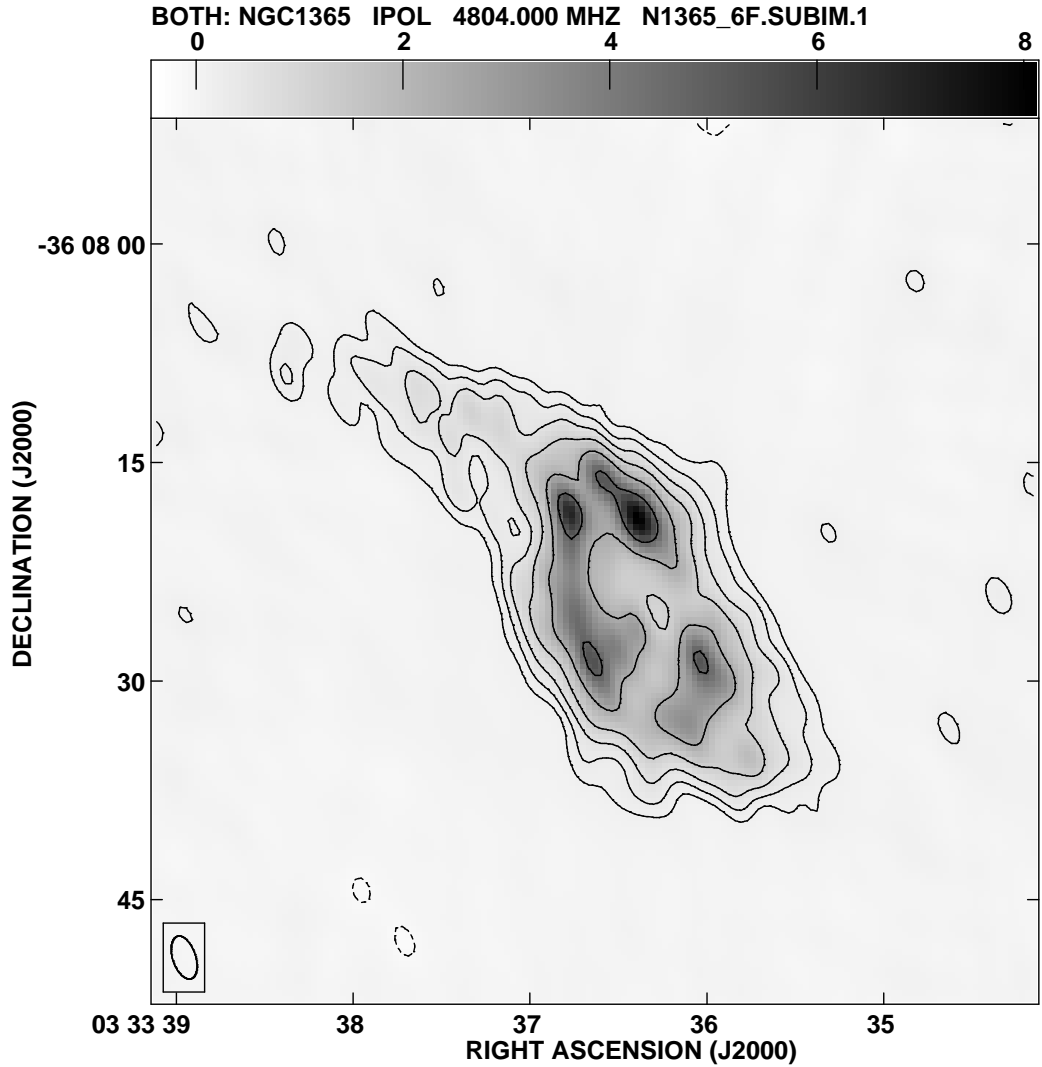,width=400pt}}
\caption{\label{fig1a}
ATCA 6cm image of NGC 1365 showing a ring of hot-spots surrounding a weak
nucleus. In this and all subsequent figures, the horizontal 
grey--scale bar above the image shows the 
grey-scale in units of mJy per beam, and contours are
at -1, 1, 2, 4, 8, 16, 32, and 64 times the lowest contour level. 
In this case the lowest contour level is 0.15 mJy/beam.
}
\end{figure*}

\begin{figure*}[p]
\centerline{\psfig{figure=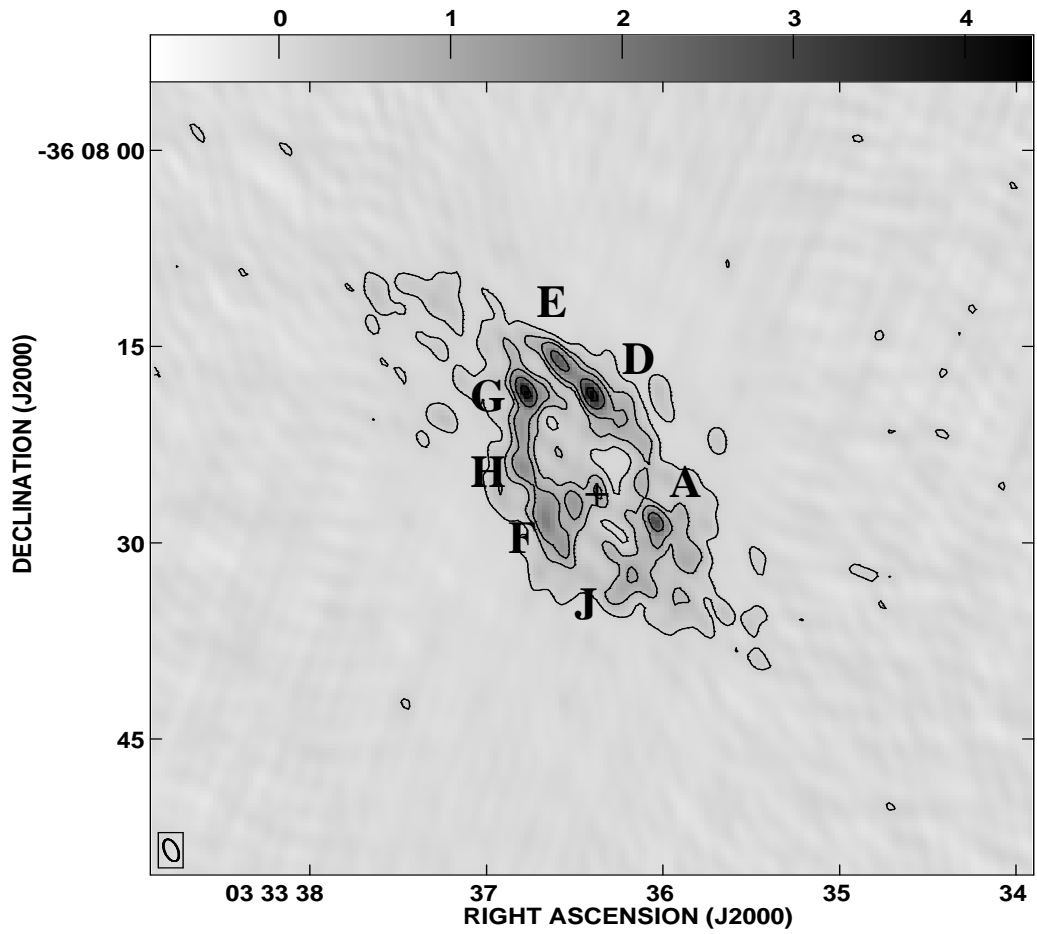,width=400pt}}
\caption{\label{fig1b}
ATCA 3cm image of NGC 1365. The lowest contour level is 0.2 mJy/beam. 
}
\end{figure*}

\begin{figure*}[p]
\centerline{\psfig{figure=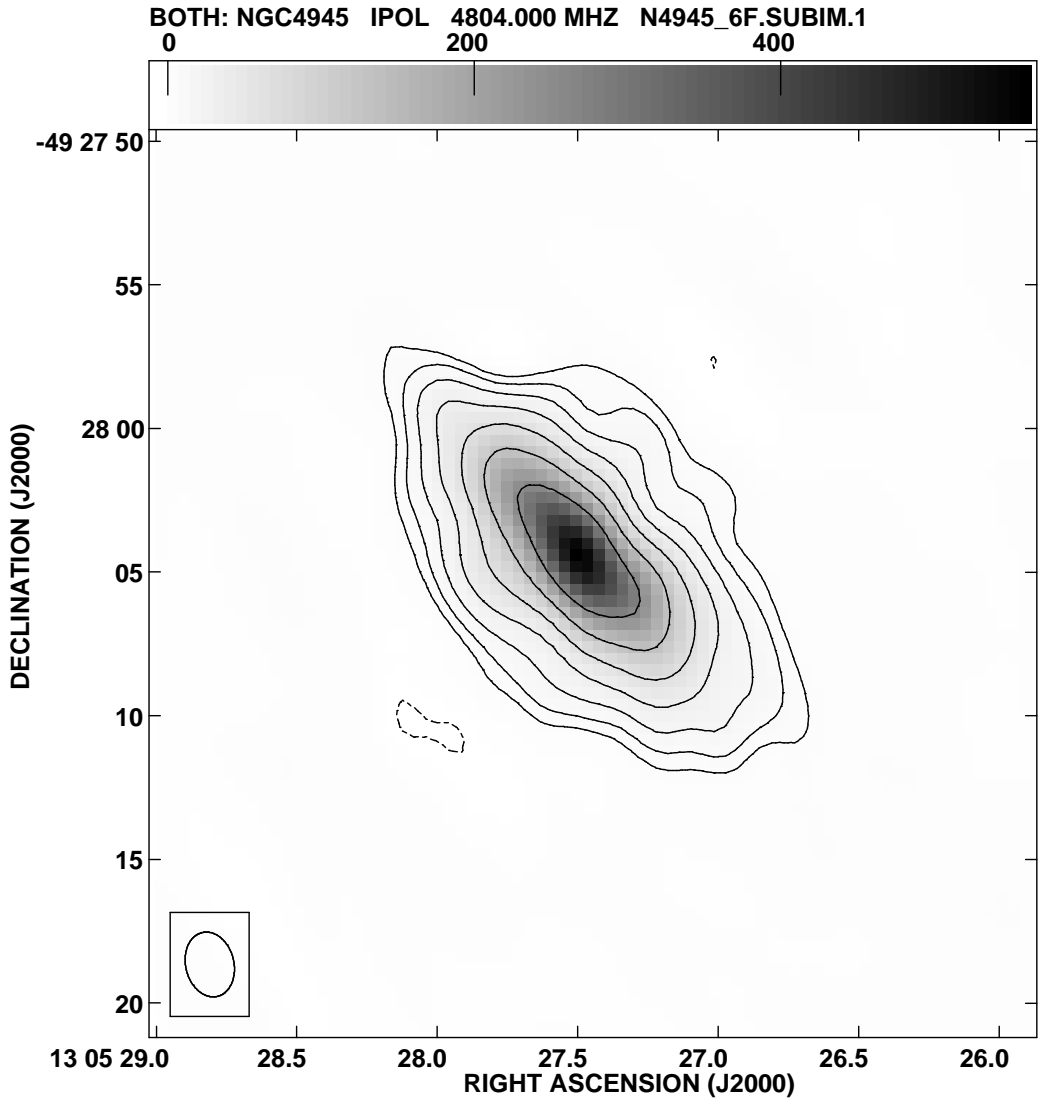,width=400pt}}
\caption{\label{fig2a}
ATCA 6cm image of NGC 4945 showing radio emission extended along the galaxy
disk with some emission perpendicular to the plane. 
The lowest contour level is 4 mJy/beam. }
\end{figure*}

\begin{figure*}[p]
\centerline{\psfig{figure=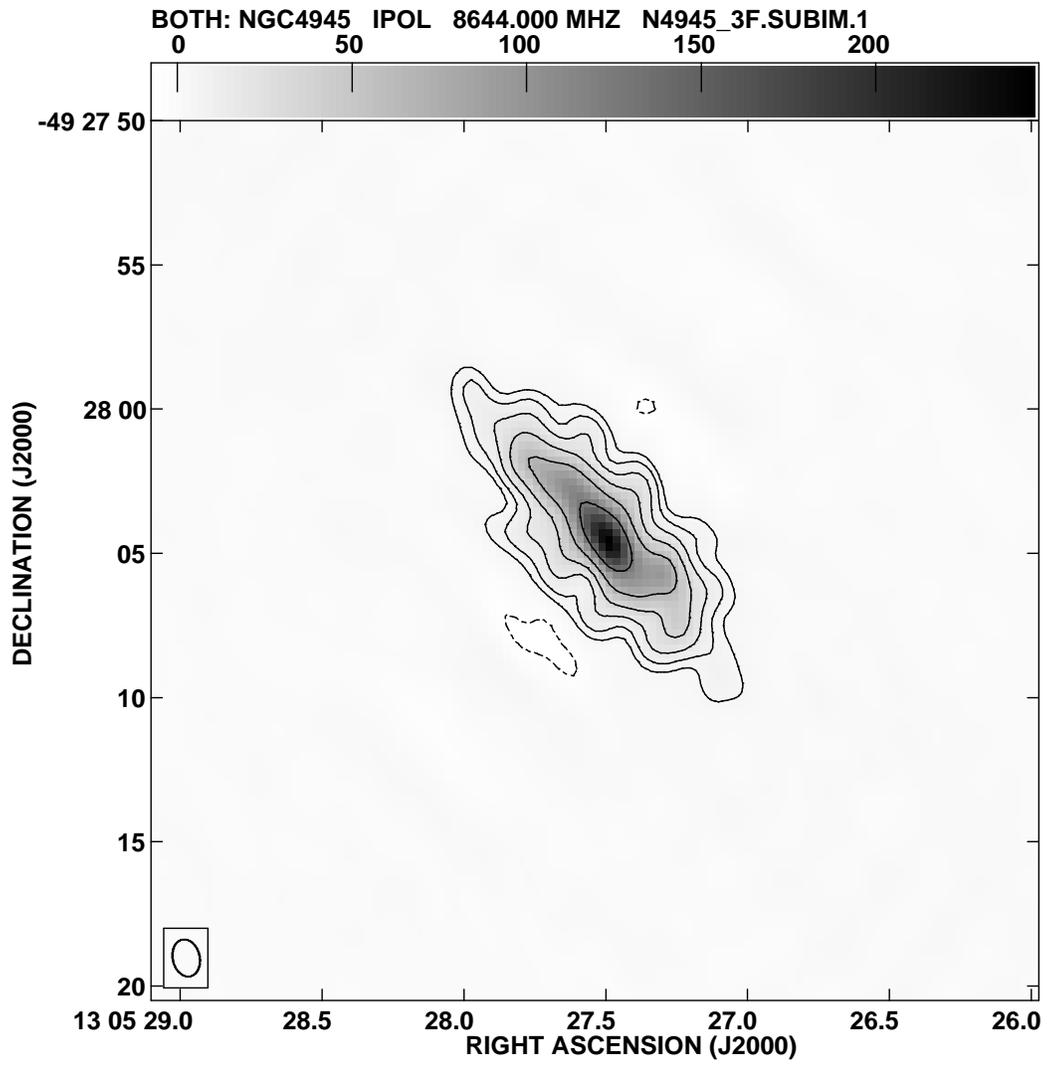,width=400pt}}
\caption{\label{fig2b}
ATCA 3cm image of NGC 4945. 
The lowest contour level is 4 mJy/beam. }
\end{figure*}

\begin{figure*}[p]
\centerline{\psfig{figure=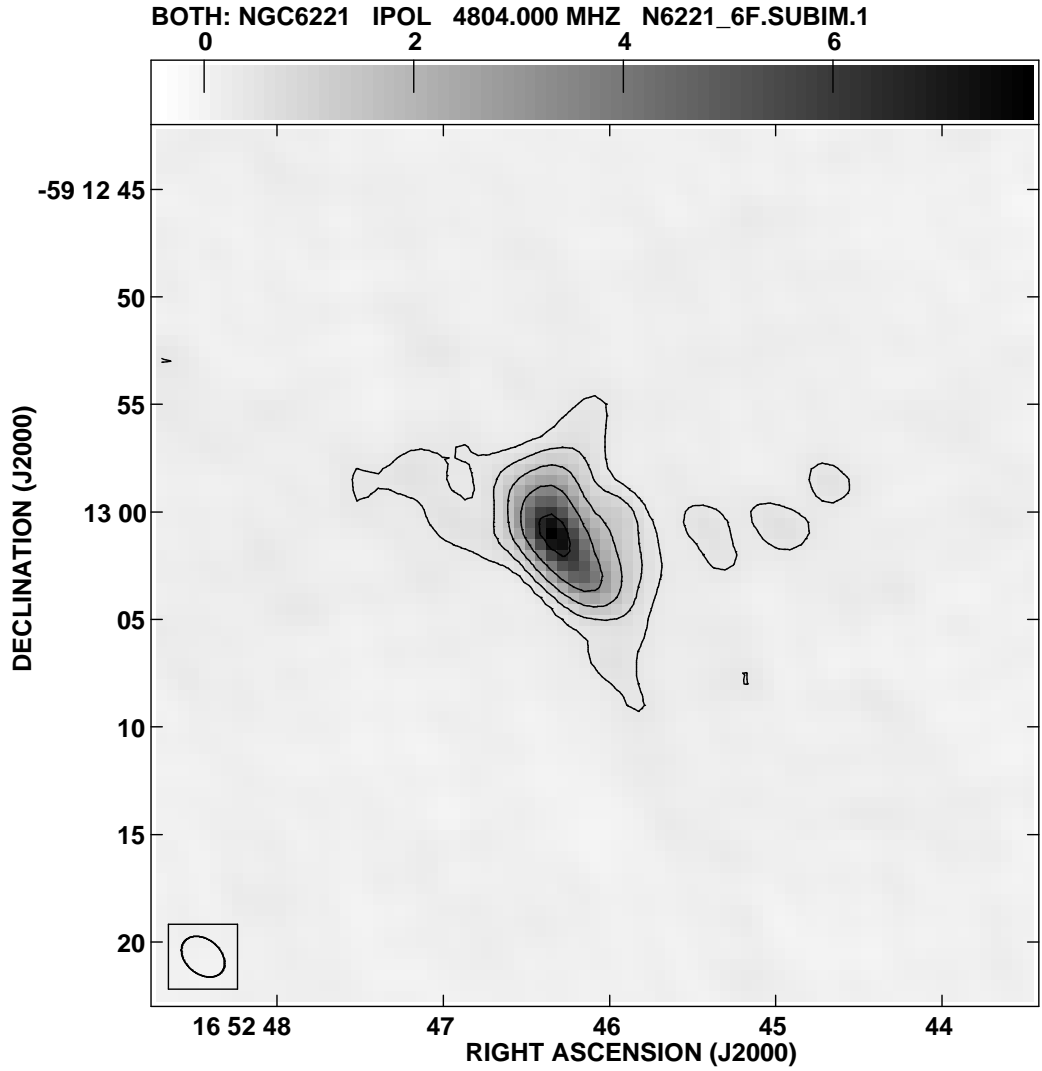,width=400pt}}
\caption{\label{fig3a}
ATCA 6cm image of NGC 6221 showing radio emission slightly extended beyond
the nucleus. 
The lowest contour level is 0.4 mJy/beam. }
\end{figure*}

\begin{figure*}[p]
\centerline{\psfig{figure=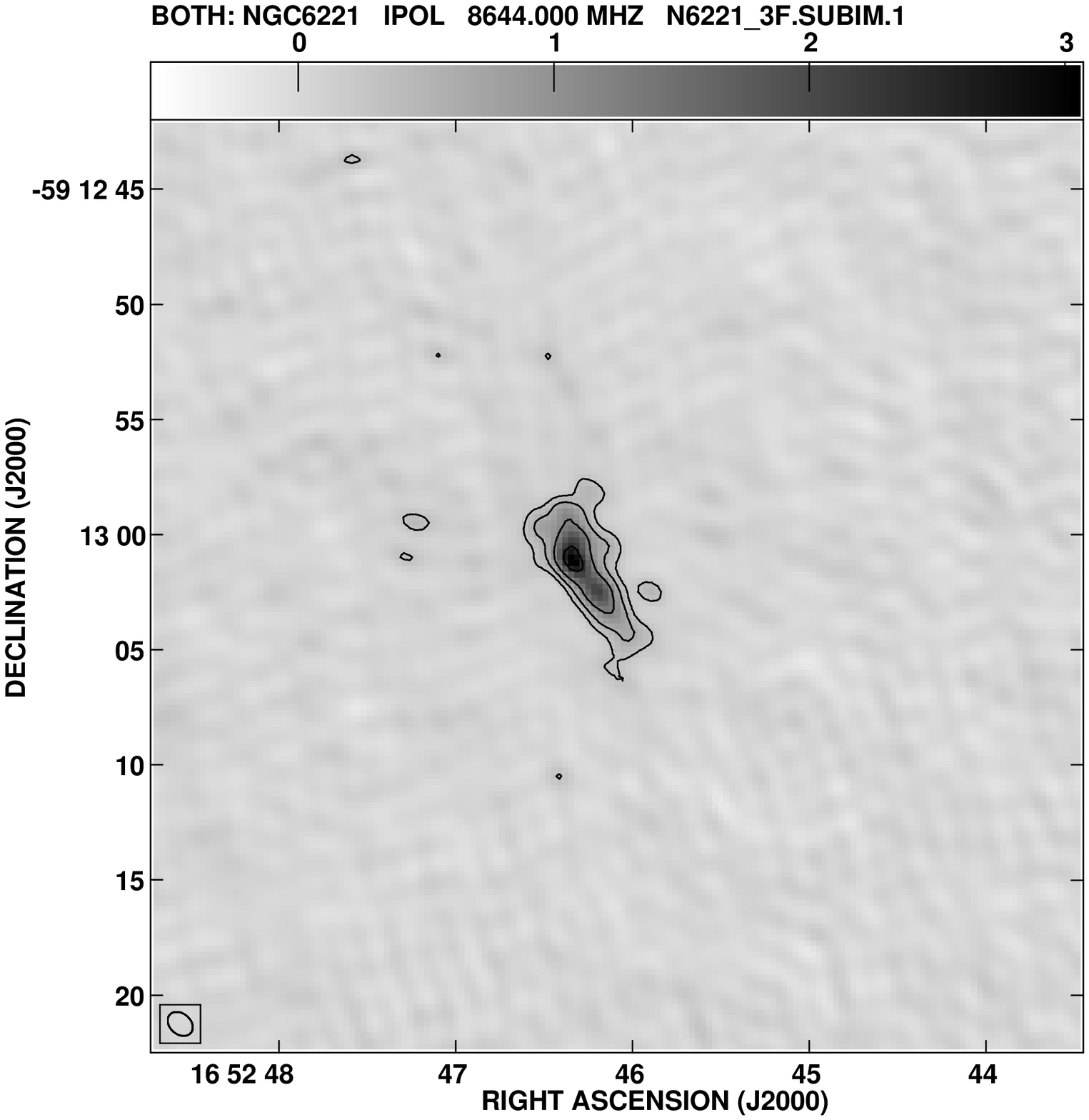,width=400pt}}
\caption{\label{fig3b}
ATCA  3cm image of NGC 6221. 
The lowest contour level is 0.3 mJy/beam. }
\end{figure*}

\begin{figure*}[p]
\centerline{\psfig{figure=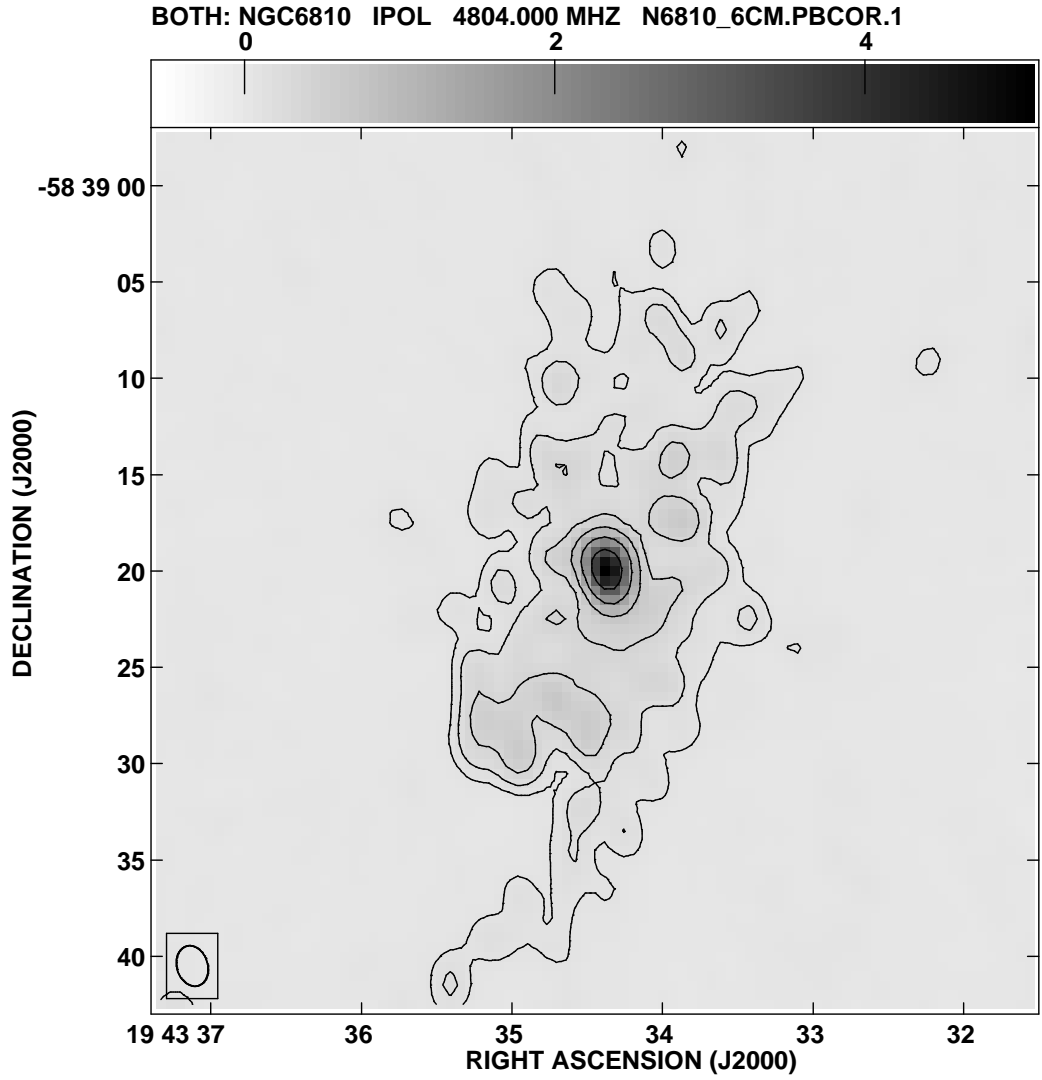,width=400pt}}
\caption{\label{fig4a}
ATCA  6cm image of NGC 6810 showing a compact nuclear source surrounded by
weak emission. The lowest contour level is 0.1 mJy/beam.
}
\end{figure*}

\begin{figure*}[p]
\centerline{\psfig{figure=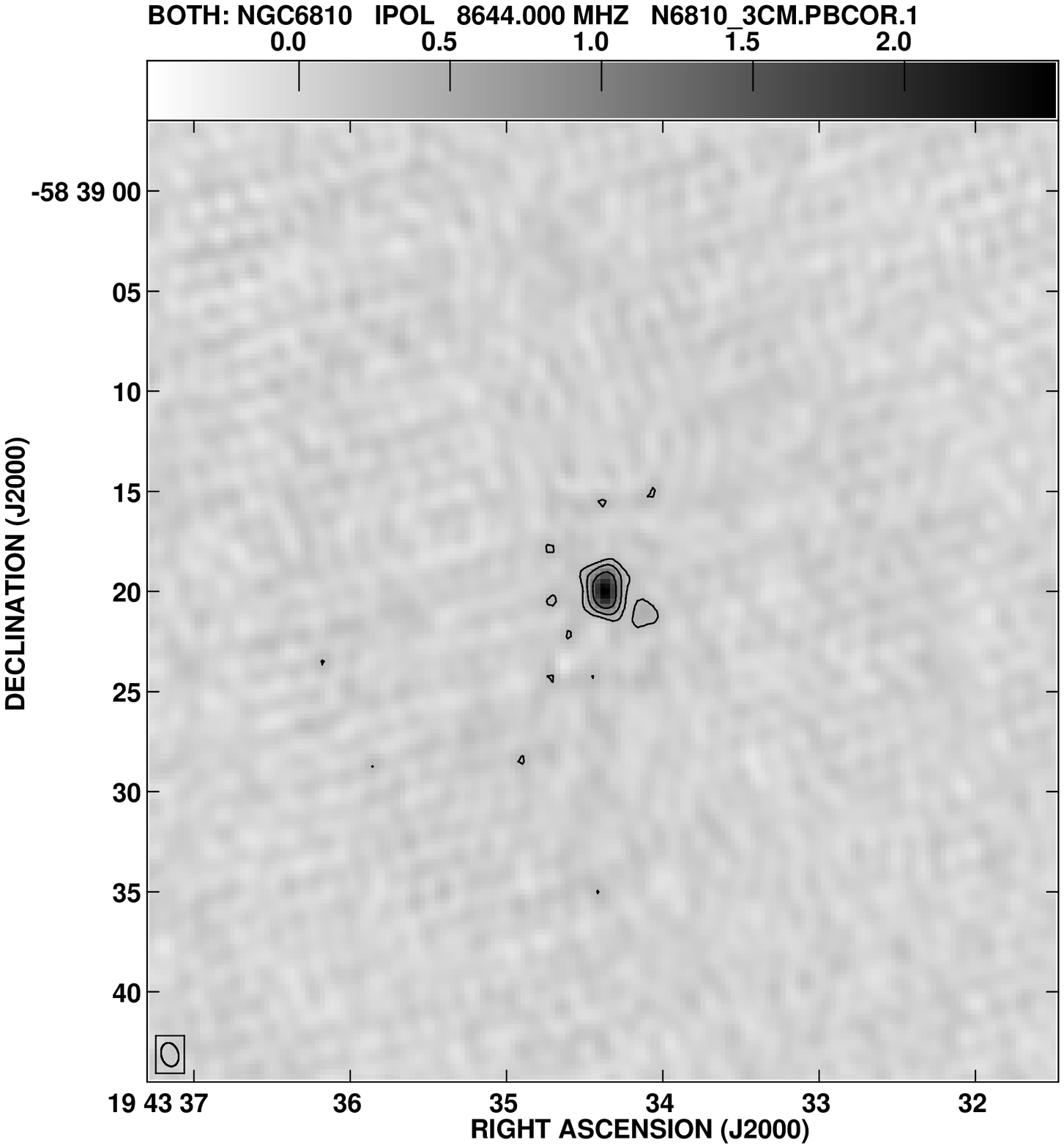,width=400pt}}
\caption{\label{fig4b}
ATCA  3cm image of NGC 6810. The lowest contour level is 0.3 mJy/beam.
}
\end{figure*}

\begin{figure*}[p]
\centerline{\psfig{figure=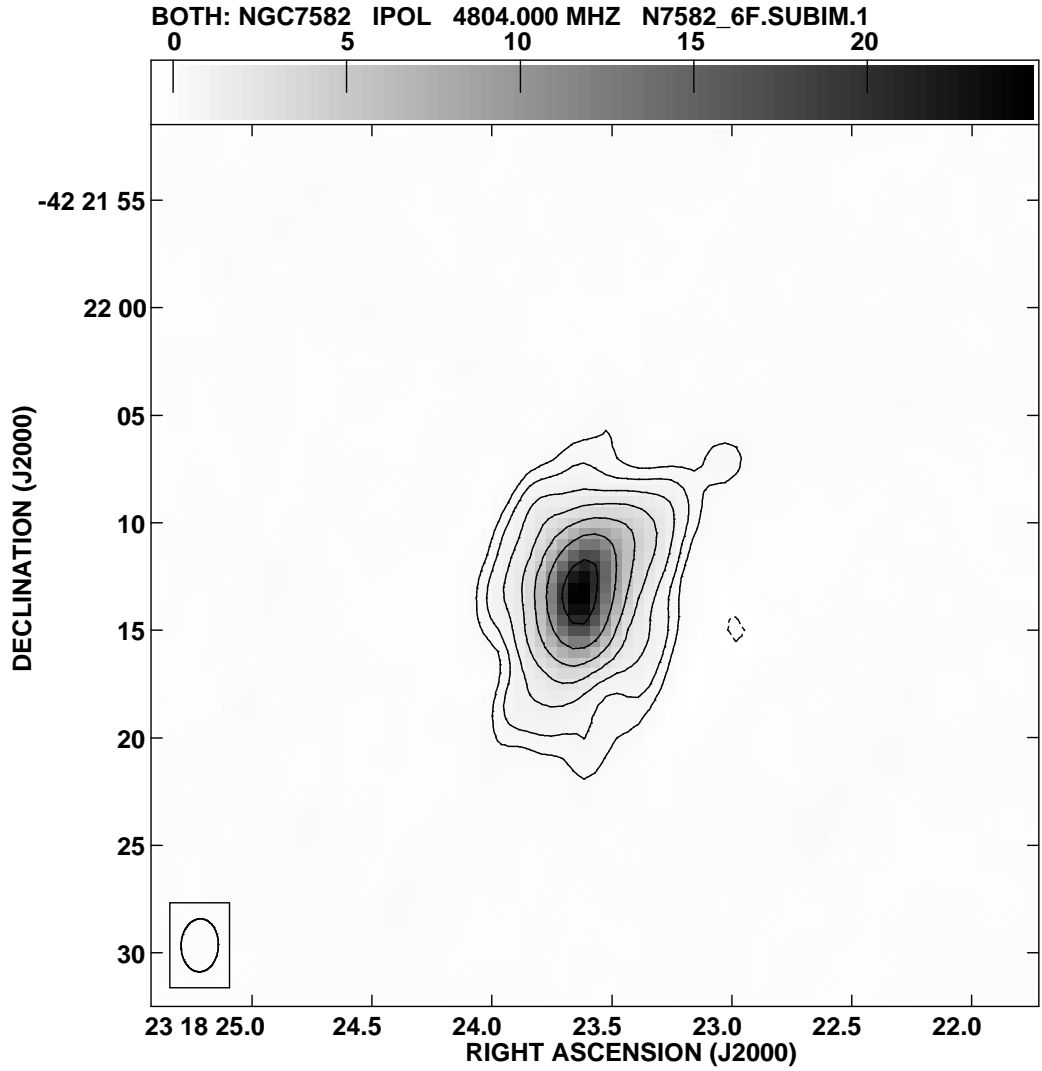,width=400pt}}
\caption{\label{fig5a}
ATCA 6cm image of NGC 7582 showing the nucleus and elongated extended 
emission. The lowest contour level is 0.2 mJy/beam.
}
\end{figure*}

\begin{figure*}[p]
\centerline{\psfig{figure=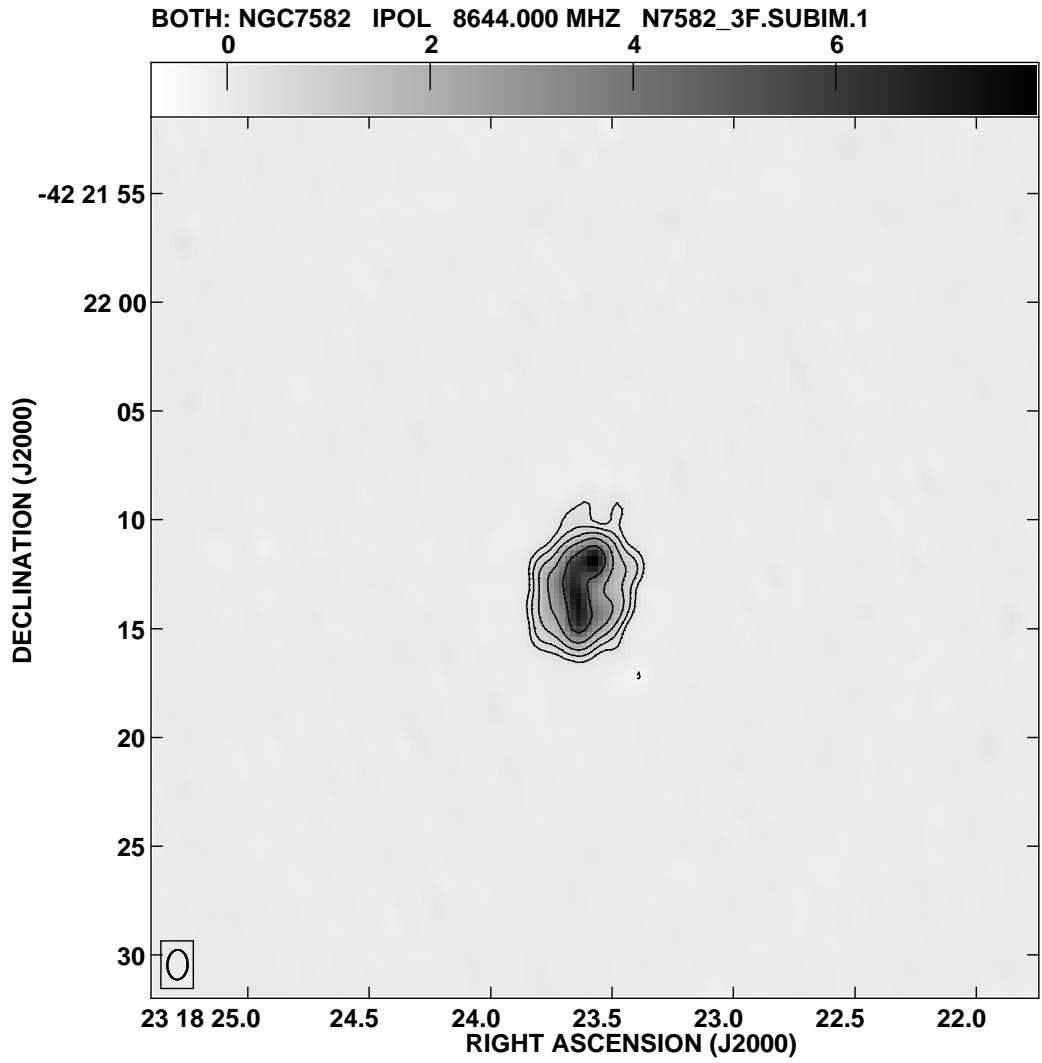,width=400pt}}
\caption{\label{fig5b}
ATCA 3cm image of NGC7582. The lowest contour level is 0.2 mJy/beam.
}
\end{figure*}

\begin{figure*}[p]
\centerline{\psfig{figure=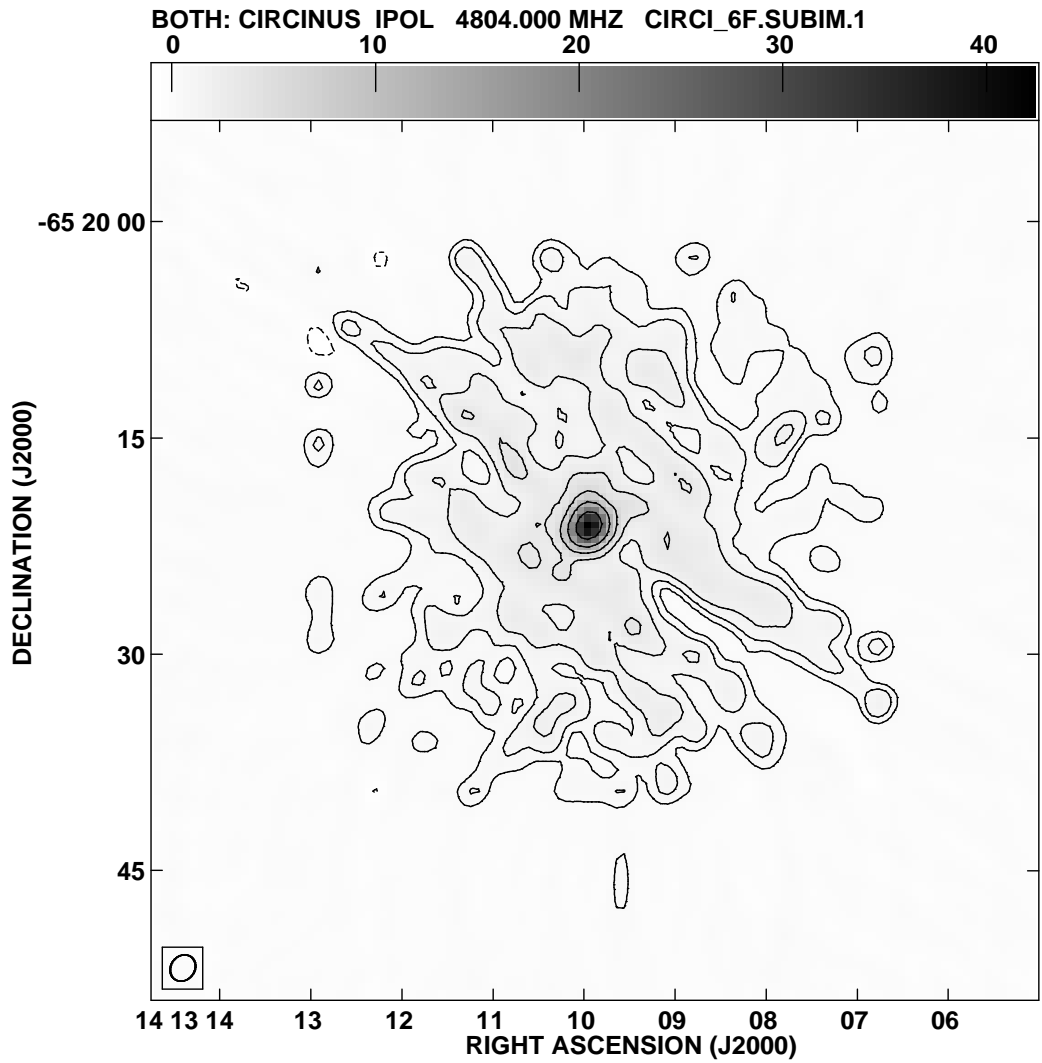,width=400pt}}
\caption{\label{fig6a}
ATCA 6cm image of Circinus showing a compact nuclear source and weak
extended emission. The lowest contour level is 0.4 mJy/beam.
}
\end{figure*}

\begin{figure*}[p]
\centerline{\psfig{figure=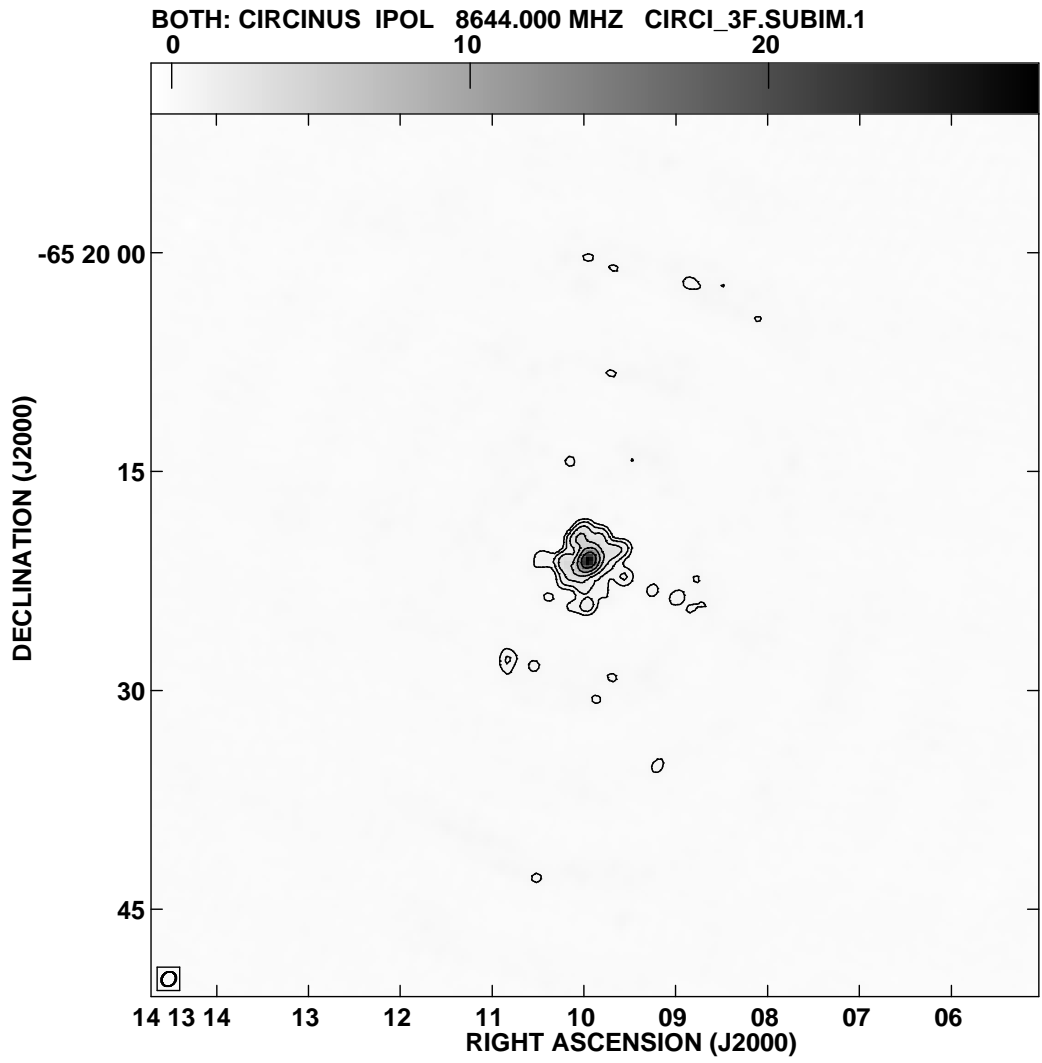,width=400pt}}
\caption{\label{fig6b}
ATCA 3cm image of Circinus. The lowest contour level is 0.5 mJy/beam.
}
\end{figure*}

\begin{figure*}[p]
\centerline{\psfig{figure=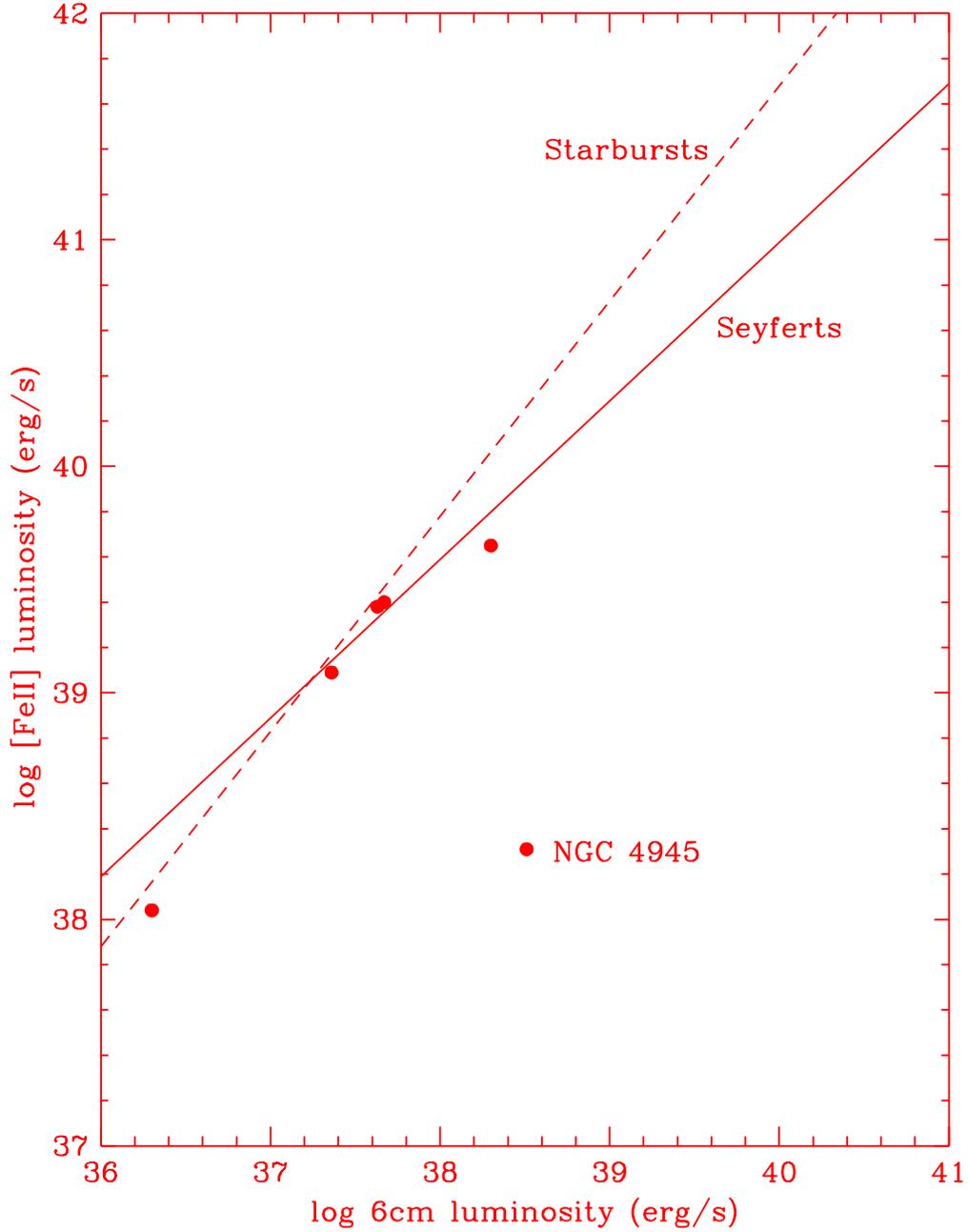,width=400pt}}
\caption{\label{fig7}
The [FeII] -- 6cm radio correlation. The dashed line represents
starburst galaxies and the solid line Seyferts, from Simpson \etal
(1996). The dispersion about each line is $\sim$ 0.5 in the log. Our
sample galaxies are shown as filled circles.
}
\end{figure*}

\centerline{\psfig{figure=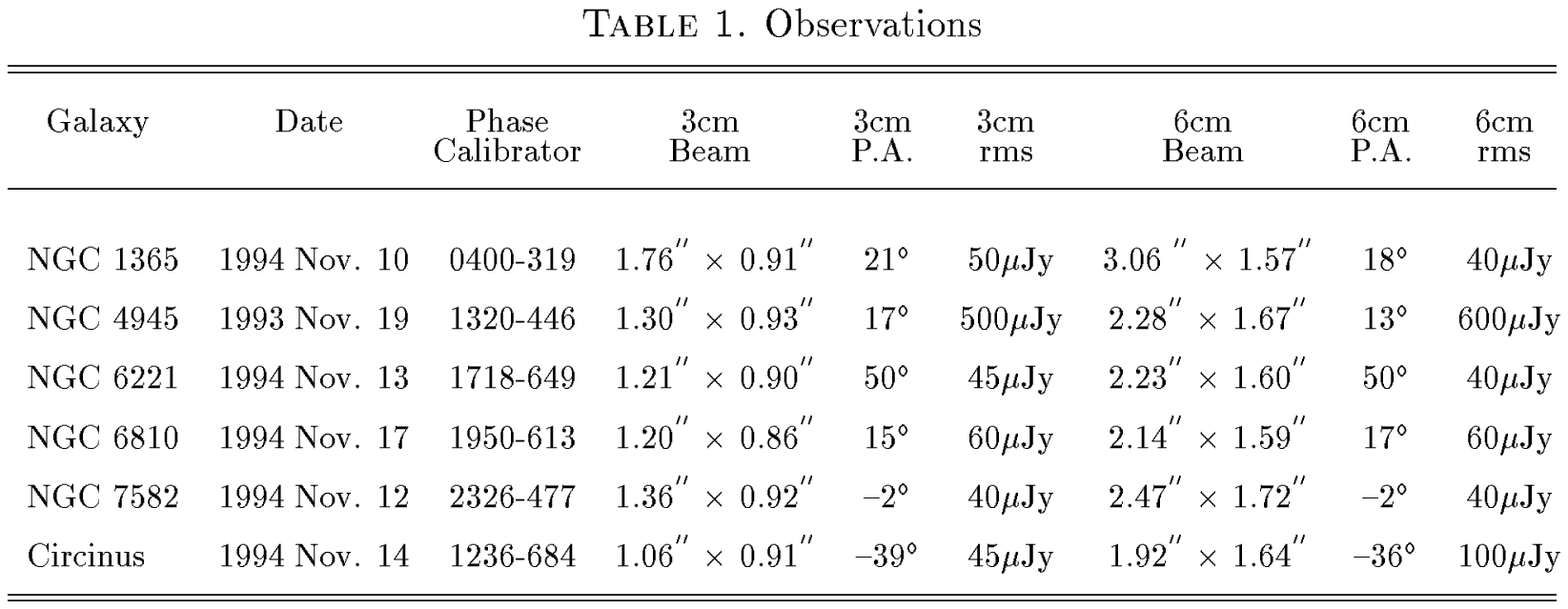,width=500pt}}

\centerline{\psfig{figure=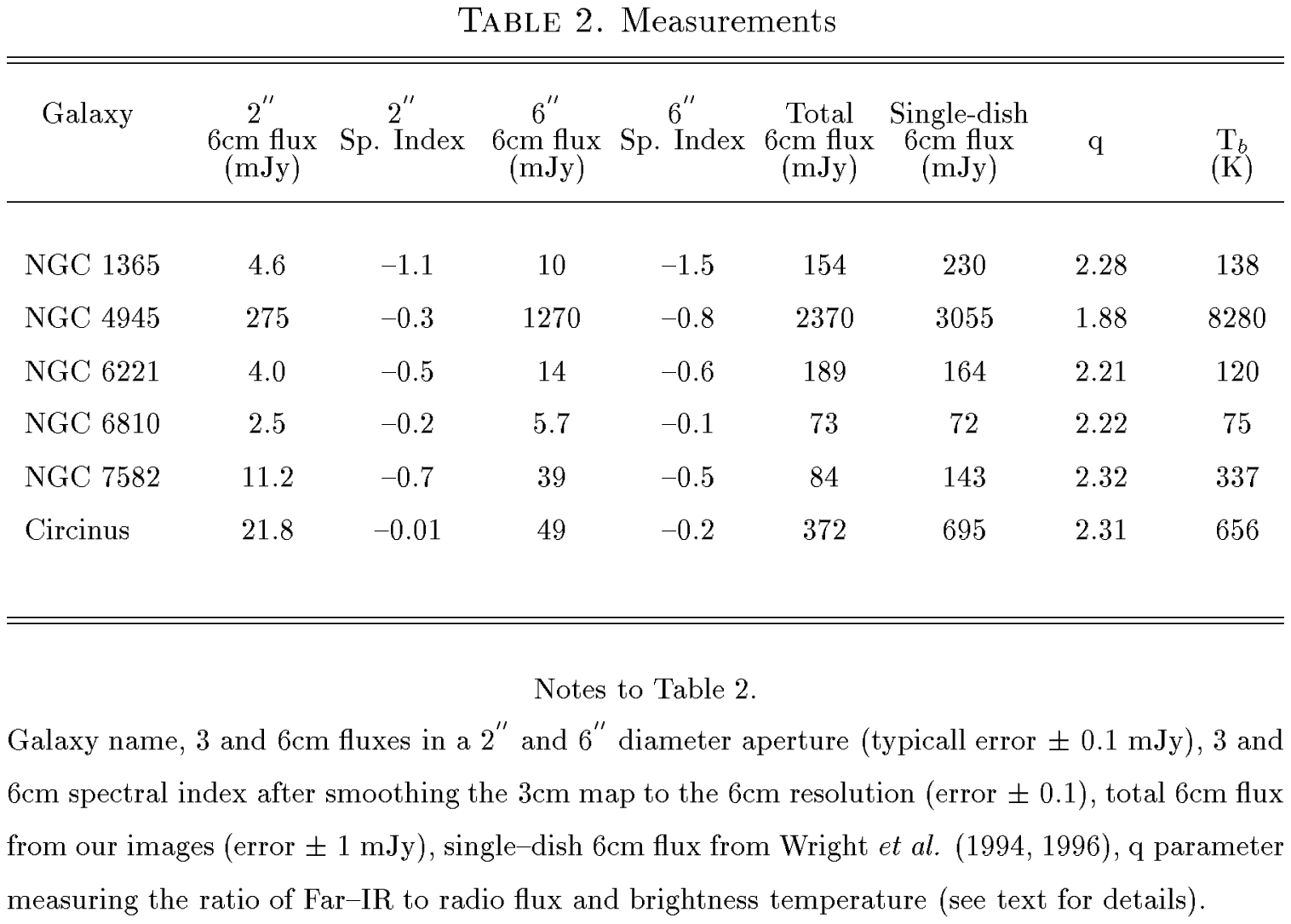,width=500pt}}

\centerline{\psfig{figure=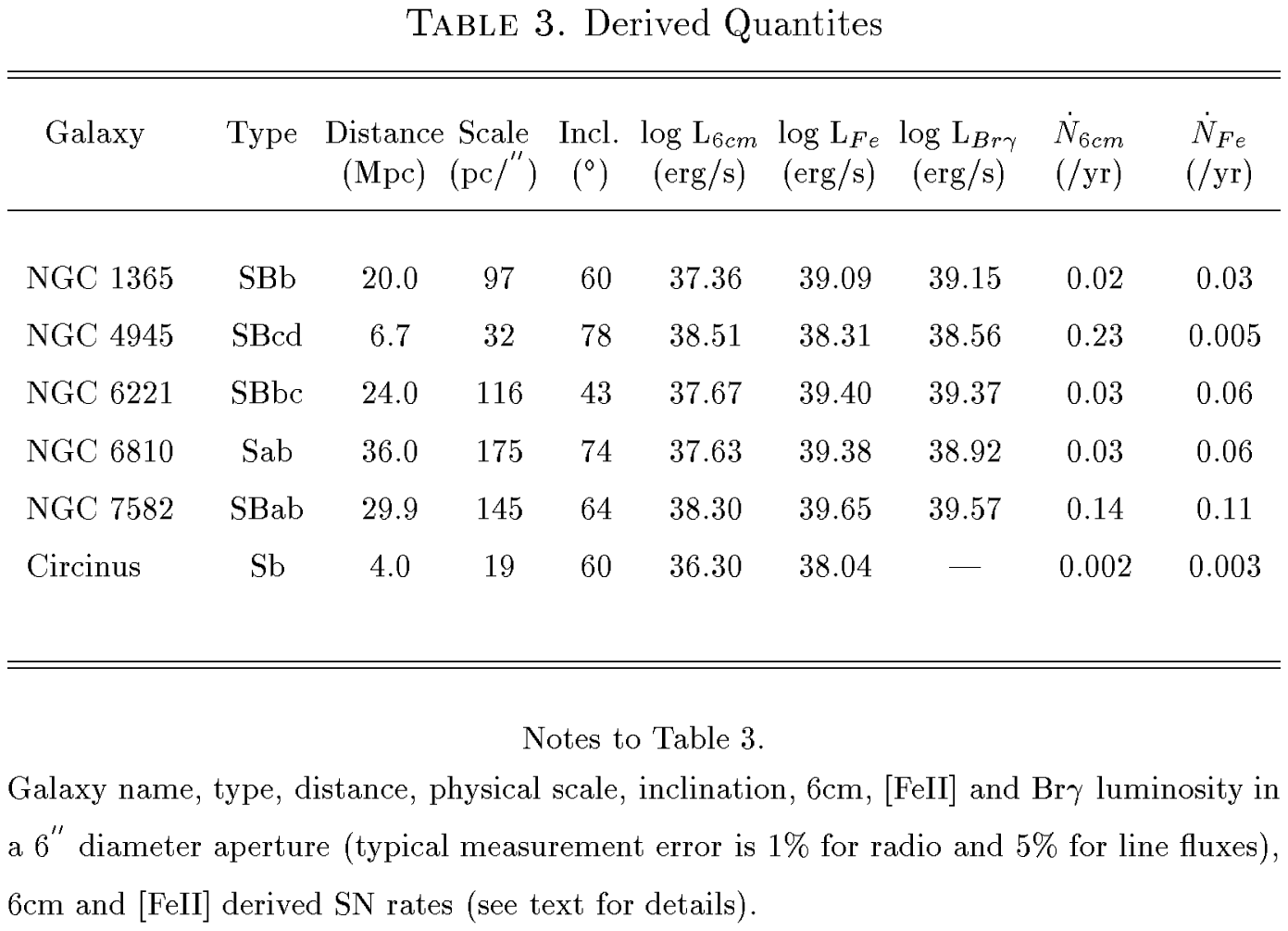,width=500pt}}


\end{document}